\documentclass{elsart}

\usepackage{epsfig}

\usepackage{amssymb}

\begin{document}

\begin{frontmatter}

\title{Effects of Neutron Irradiation on Carbon Doped MgB$_2$ Wire Segments}

\author{R. H. T. Wilke, S. L. Bud'ko, P. C. Canfield, and D. K. Finnemore}

\address{Ames Laboratory US DOE and Department of Physics and Astronomy, Iowa State University, Ames, IA 50011}

\author{Raymond J. Suplinskas$^\dagger$}

\address{Specialty Materials, Inc., 1449 Middlesex Street, Lowell, Massachusetts 01851}

\author{J. Farmer}
\address{Missouri University Research Reactor, University of
Missouri - Columbia, Research Park, Columbia, MO 65211}

\author{S. T. Hannahs}

\address{National High Magnetic Field Laboratory, Florida State University, 1800 E. Paul Dirac Drive, Tallahassee, Florida 32310}

\begin{abstract}
We have studied the evolution of superconducting and normal state
properties of neutron irradiated Mg(B$_{.962}$C$_{.038}$)$_2$ wire
segments as a function of post exposure annealing time and
temperature. The initial fluence fully suppressed
superconductivity and resulted in an anisotropic expansion of the
unit cell. Superconductivity was restored by post-exposure
annealing. The upper critical field, H$_{c2}$(T=0), approximately
scales with T$_c$ starting with an undamaged T$_c$ near 37 K and
H$_{c2}$(T=0) near 32 T. Up to an annealing temperature of 400 $^
o$C the recovery of T$_c$ tends to coincide with a decrease in the
normal state resistivity and a systematic recovery of the lattice
parameters. Above 400 $^ o$C a decrease in order along the c-
direction coincides with an increase in resistivity, but no
apparent change in the evolution of T$_c$ and H$_{c2}$. To first
order, it appears that carbon doping and neutron damaging effect
the superconducting properties of MgB$_2$ independently.
\end{abstract}
\end{frontmatter}

\section{Introduction}

Neutron irradiation and carbon doping of MgB$_2$ are two methods
by which defects can systematically be introduced. Carbon enters
the structure replacing boron \cite{1,2} and electron dopes the
system, which results in a suppression of T$_c$ \cite{3}. Carbon
presumably acts as a point defect, enhancing scattering primarily
within the $\pi$ band \cite{3} which leads to an enhancement of
H$_{c2}$ at low doping levels \cite{4} in accordance with
theoretical calculations \cite{5}. Up to a doping level of 10\%,
two gap superconductivity is preserved \cite{6} and evidence
suggests that it may persist down to T$_c$=0 \cite{7}.

Due to the high neutron capture cross section of $^{10}$B, neutron
irradiation studies have varied as researchers have employed
different techniques to ensure uniform damage throughout samples.
The superconducting properties differ depending upon the
irradiation conditions. Enhancements in H$_{c2}$ have been
observed for neutron irradiation on isotopically enriched
Mg$^{11}$B$_2$ for fluences of order 10$^{17}$ n/cm$^2$ \cite{14},
and for irradiation of MgB$_2$ containing natural boron with a
fast neutron fluence of 10$^{19}$ n/cm$^2$ \cite{15}. Defects
introduced by neutron irradiation, regardless of irradiation
conditions, are effective at pinning vortices and for samples
which exhibit a T$_c$ above approximately 30 K show enhanced
J$_c$(H) values in the vicinity of 20 K \cite{14,16,17}.

We have previously reported on neutron irradiation of MgB$_2$
filaments \cite{8}. Although the filaments contained natural boron
and MgB$_2$ containing natural boron has a slow neutron half depth
of 130 $\mu$m, they were 140 $\mu$m in diameter and, given that an
isotropic neutron flux was used, the damage is believed to be
essentially uniform throughout the samples. Irradiating with a
fluence level of 4.75*10$^{18}$ n/cm$^2$ resulted in a suppression
of T$_c$ to below 5 K. Superconductivity was restored by
performing post exposure anneals. Upper critical field values were
found to approximately scale with T$_c$. It is believed that the
suppression of T$_c$ is a result of both an increase in interband
scattering and a decrease in the electron density of states at the
Fermi surface. The temperature dependance of the upper critical
field exhibits Werthamer, Helfand, and Hohenberg (WHH) \cite{13}
like behavior, which suggests the bands become fully mixed, only
when T$_c$ is near or below 10 K. The field dependence of the
critical current density was found to depend largely upon the
superconducting transition temperature. Generally speaking,
J$_c$(H,T) increases for samples with higher T$_c$ values. For
example the field at which filaments could carry in excess of
10$^4$ A/cm$^2$ was extended from near 1 T in an undamaged wire to
above 1.5 T in the case of a sample whose annealing profile
resulted in a T$_c$ near 37 K.

Given that carbon doping appears to weakly decrease T$_c$ and
increase H$_{c2}$ by preferentially increasing scattering within
the $\pi$ band, whereas neutron damaging appears to decrease T$_c$
and H$_{c2}$ by increasing the inter-band scattering, the focus of
this paper is to examine the effects of combining these two
scattering mechanisms by inducing and removing the effects of
neutron damage in carbon doped MgB$_2$ samples. This will allow us
to add and subtract inter-band scattering in samples with already
enhanced intra-$\pi$-band scattering.

\section{Experimental Methods}

Carbon doped, Mg(B$_{.962}$C$_{.038}$)$_2$, was prepared in a two
step reaction process as described in detail elsewhere \cite{9}.
Carbon doped boron filaments, produced by Specialty Materials,
Inc., were exposed to excess Mg vapor while the temperature was
ramped from 650 $^o$C to 1200 $^o$C over 96 hours. Three
filaments, each approximately 5 mm in length, were sealed in
quartz ampoules under a He atmosphere and irradiated with an
isotropic fluence of 7.13$*$10$^{18}$ n/cm$^2$ reactor neutrons at
the Missouri University Research Reactor (MURR), as described in
reference \cite{8}. The Mg(B$_{.962}$C$_{.038}$)$_2$ filaments had
a diameter of 110 $\mu$m, indicating the isotropic irradiation
should result in essentially uniform damage. The boron filaments
used in this study contain a tungsten boride core. Fast neutrons
colliding with $^{182}$W atoms, which have a natural abundance of
26.3\%, can be absorbed into the nucleus causing the emission of a
proton and transforming the tungsten into $^{182}$Ta. $^{182}$Ta
$\beta$ decays back to $^{182}$W, with a half life of 181 days. As
a result the filaments were mildly radioactive and required
appropriate safety measures in handling.

Normal state and superconducting properties were determined for a
series of post exposure annealing profiles. One set consisted of
24 hour anneals at temperatures up to 600 $^o$C. In the second set
the annealing temperature was held constant at 500 $^o$C while the
annealing time was varied from 1-1000 hours. In all cases the
anneals were performed by placing samples, still sealed with
their quartz ampoules, into a Lindberg model 55035 Mini-Mite tube
furnace that was preheated to the desired annealing temperature.
After the samples were annealed for the desired length of time,
the ampoules were removed from the furnace and air quenched to
room temperature.

Powder x-ray diffraction (XRD) measurements were made at room
temperature using CuK$\alpha$ radiation in a Rigaku Miniflex
Diffractometer. Measurements were performed on six filaments from
two ampoules. Peak positions were determined by fitting each peak
with a Pseudo-Voigt function using Jade analysis software. A
silicon standard was used to calibrate each pattern. The
experimentally determined Si peak positions were found to be
offset from their known values by a constant amount. Within each
spectra, the peaks varied about some constant offset and this
variation was used to estimate experimental uncertainty in the
lattice parameters. Lattice parameters were determined from the
position of the (002) and (110) peaks. DC magnetization
measurements and magnetization hysteresis loops were done in a
Quantum Design MPMS-5 SQUID magnetometer. For magnetization
measurements, individual wires, which have of mass of
approximately 0.15 mg, were oriented along the direction of the
applied field. Transport measurements were done using a standard
four probe AC technique, with platinum wires attached to the
samples with Epotek H20E silver epoxy. Typical samples had voltage
contacts that were 2-3 mm apart. Resistivity versus temperature in
applied fields up to 14 T were carried out in a Quantum Design
PPMS-14 system and resistivity versus field was measured up to
32.5 T using a lock-in amplifier technique at the National High
Magnetic Field Laboratory in Tallahassee, Florida.

\section{Structural Properties}

\begin{figure}
\begin{center}
\includegraphics[angle=0,width=180mm]{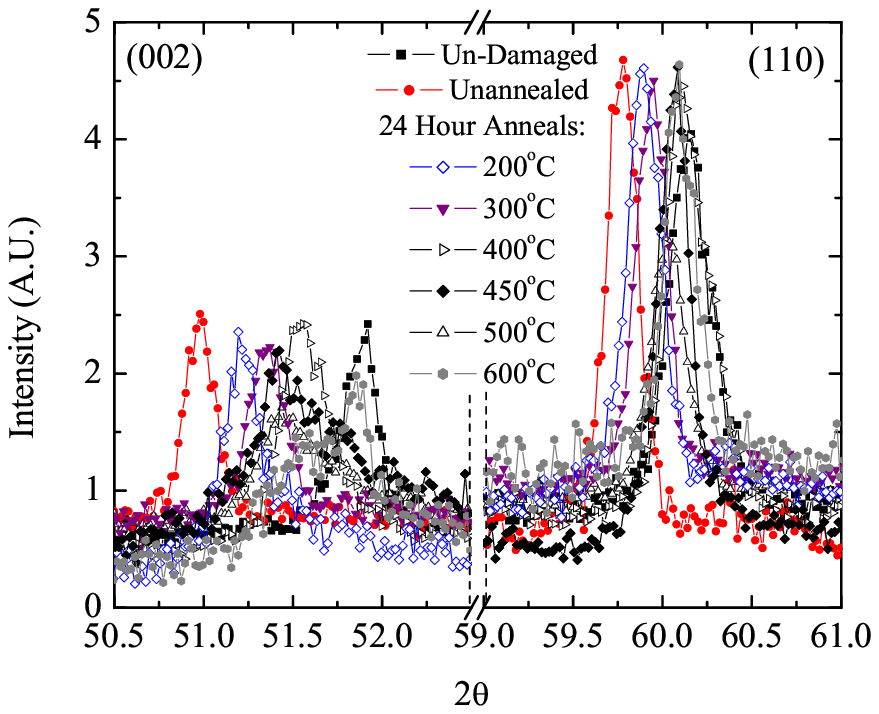}
\end{center}
\caption{(002) and (110) x-ray peaks for the series of neutron
irradiated Mg(B$_{.962}$C$_{.038}$)$_2$ samples annealed for 24
hours at various temperatures. }\label{24hpeaks}
\end{figure}

Figure \ref{24hpeaks} presents the (002) and (110) x-ray peaks for
the entire set of 24 hour anneals. In the as-damaged samples both
peaks are shifted to lower 2$\theta$, indicating the irradiation
resulted in an expansion of the unit cell. The calculated a- and
c- lattice parameters yield relative increases with respect to the
undamaged sample of $\Delta$a=0.0168(7) $\AA$ and
$\Delta$c=0.0650(10) $\AA$. In the case of neutron irradiation on
pure MgB$_2$, samples exposed to a fluence of 4.75*10$^{18}$
n/cm$^2$ showed an increase in the a- and c- lattice parameters of
0.0113(7) $\AA$ and 0.0538(9) $\AA$. At a higher fluence of
9.50*10$^{18}$ n/cm$^2$ a greater expansion of the lattice
parameters was observed, with a- and c- increasing by 0.0141(9)
$\AA$ and 0.0596(12) $\AA$, respectively \cite{8}. Thus the
magnitude of the increase in the lattice parameters is greater in
the case of the carbon doped MgB$_2$ samples than was observed in
pure MgB$_2$ samples even when they are exposed to a higher
fluence level.

\begin{figure}
\begin{center}
\includegraphics[angle=0,width=180mm]{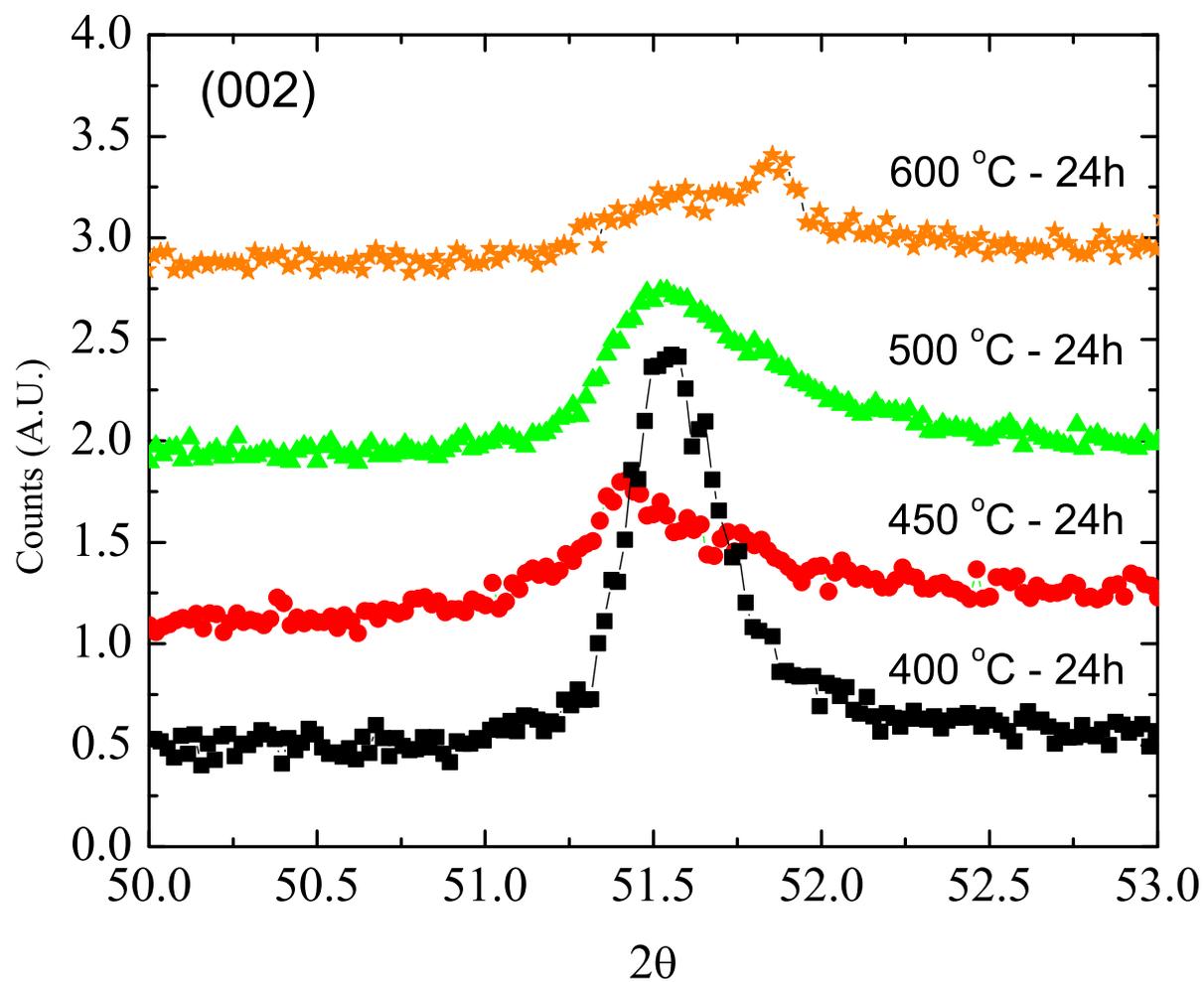}
\end{center}
\caption{Evolution of the x-ray (002) peak as a function of
annealing temperature for 24 hour anneals.}\label{xray002peak}
\end{figure}
\begin{figure}
\begin{center}
\includegraphics[angle=0,width=180mm]{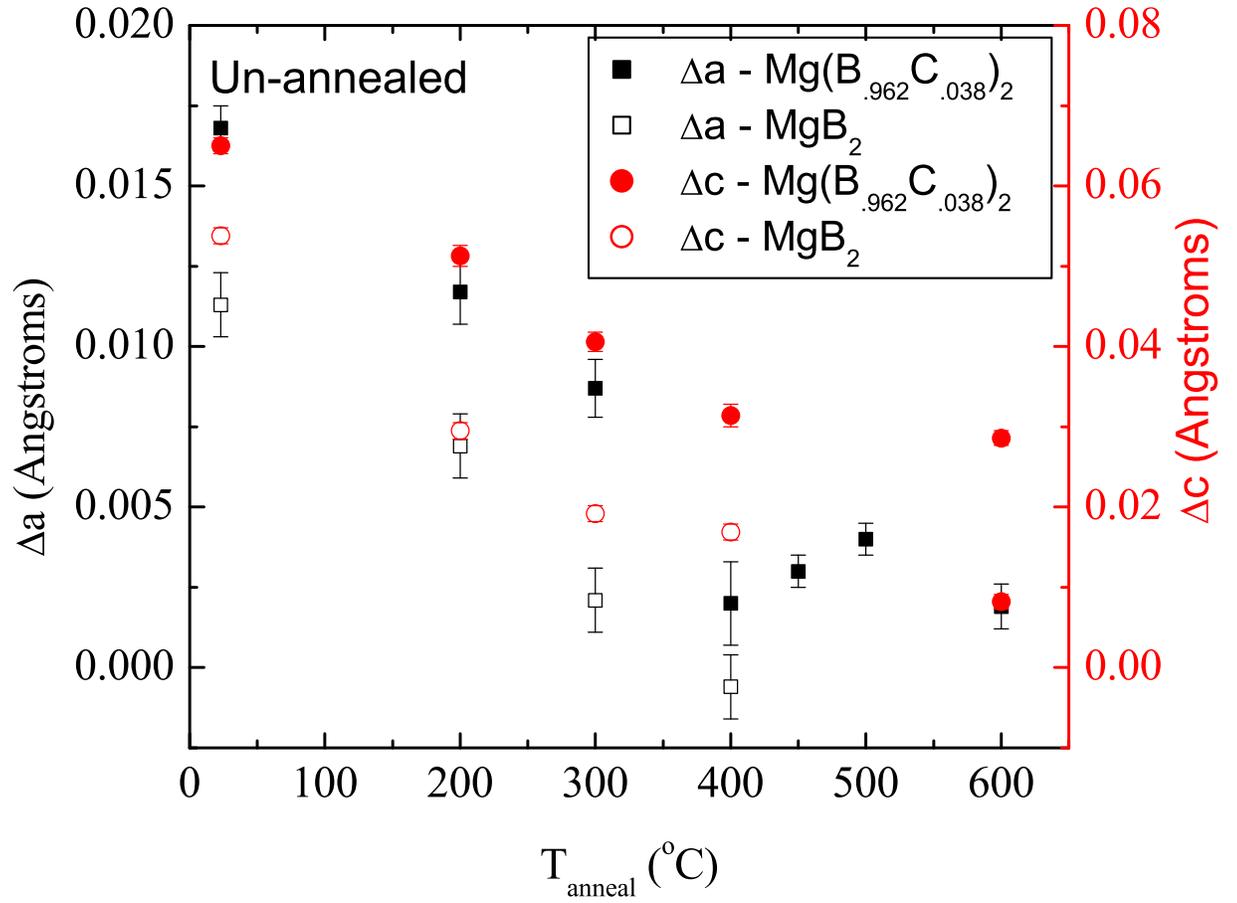}
\end{center}
\caption{Lattice parameters as a function of annealing temperature
for 24 hour anneals, as calculated from the positions of the (002)
and (110) x-ray peaks. Included are data on a pure MgB$_2$ sample
exposed to a fluence of 4.75*10$^{18}$ n/cm$^2$ from reference
\cite{8}. }\label{deltaparameters}
\end{figure}

As the annealing temperature is increased up to a temperature of
400 $^o$C we see a systematic shift of both the (002) and (110)
peaks to higher 2$\theta$, indicating a contraction of both the a-
and c- lattice parameters. At 400 $^o$C the a-lattice parameter
appears to reach a minimum, measuring 3.0769(11) $\AA$, which is
0.0020(13) $\AA$ or 0.065\% larger than the undamaged sample. The
c-lattice parameter decreases monotonically as a function of
temperature up to 400 $^o$C, where it has a value of 3.5466(12)
$\AA$, which is 0.0314(14) $\AA$ or 0.89\% larger than the
undamaged value. For anneals at 450 $^o$C and above, a qualitative
change in the evolution of the x-ray peaks occurs. The (002) peaks
begin to broaden substantially (Figure \ref{xray002peak}). For the
450 $^o$C and 500 $^o$C anneals, the stable peak refinements for
the (002) peak were unattainable, preventing us from attaining
estimates of the c- lattice parameter. The corresponding (110)
peaks shifts to lower 2$\theta$, indicating that the a- lattice
parameter may be expanding slightly. After annealing at 600 $^o$C
for 24 hours, the (002) peak appears to bifurcate, and indexing as
two different peaks yields one $\Delta$c value which is comparable
to that attained for an annealing at 400 $^o$C and another which
is considerably lower (Figure \ref{deltaparameters}). The (110)
peak position, and hence calculated $\Delta$a, is comparable to
that of the 400 $^o$C anneal. The full set of calculated changes
in lattice parameters relative to the undamaged samples for the
series of 24 hour anneals is plotted in figure
\ref{deltaparameters}. Included in figure \ref{deltaparameters} is
the evolution of the relative change in the lattice parameters as
a function of annealing temperature for a pure MgB$_2$ sample
exposed to a fluence of 4.75*10$^{18}$ n/cm$^2$ from reference
\cite{8}.

\begin{figure}
\begin{center}
\includegraphics[angle=0,width=180mm]{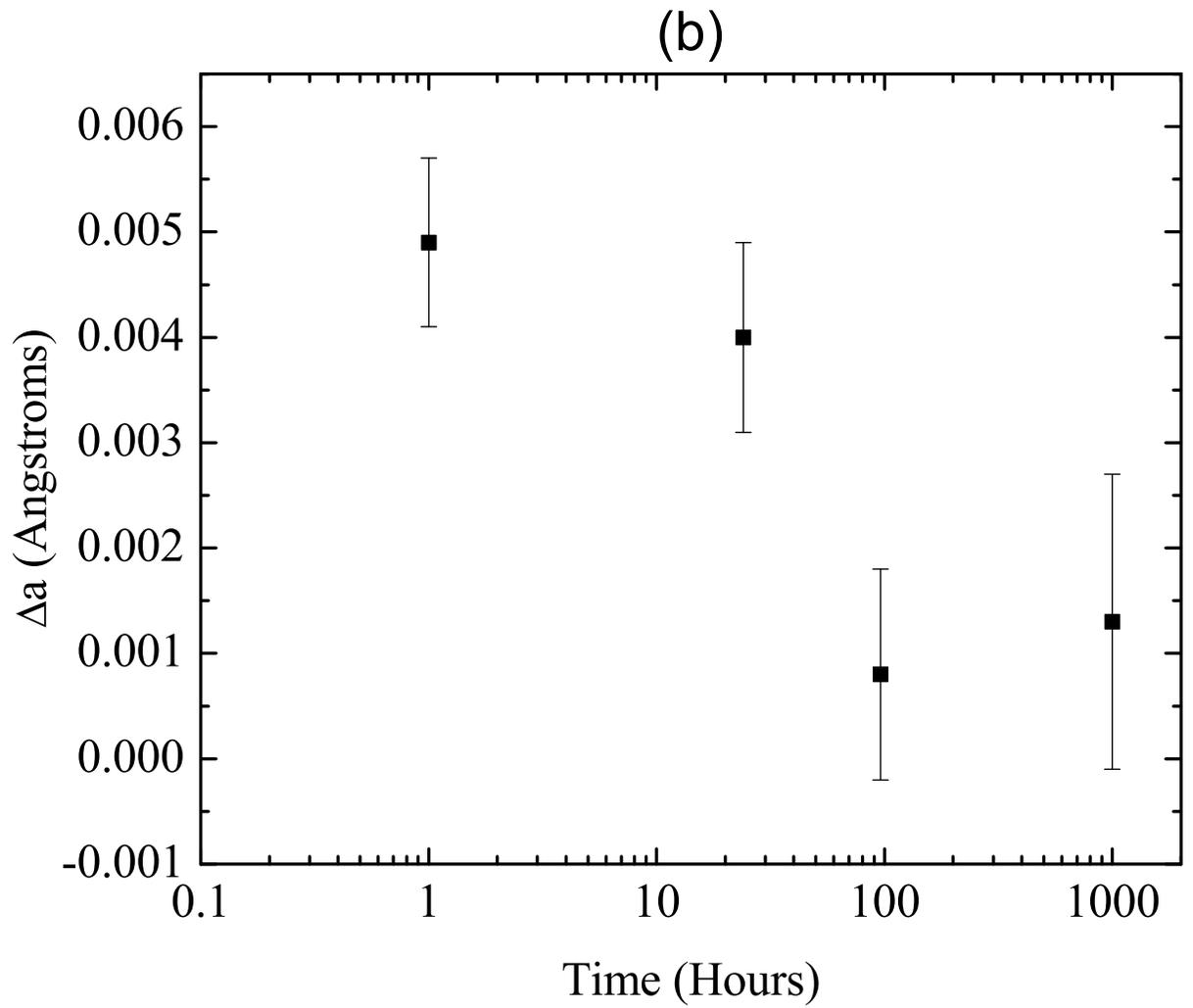}
\end{center}
\caption{(a) Evolution of the (002) and (110) x-ray peaks as a
function of annealing time at 500 $^o$C. (b) Calculated shift of
the a- lattice parameter from the undamaged sample.
}\label{anneal500c}
\end{figure}

The (002) and (110) x-ray peaks for a series of samples annealed
at 500 $^o$C for times up to 1000 hours are shown in figure
\ref{anneal500c}a. The (002) peaks continue to broaden as the
annealing time at 500$^o$C is increased. Stable peak refinements
for the (002) peaks were unattainable due to their distorted
shape. The (110) peaks monotonically shift to higher 2$\theta$ as
a function of annealing time up to 96 hours, at which point the a-
lattice parameter is within experimental error of the undamaged
value (Figure \ref{anneal500c}b). While the a- lattice parameter
is larger for the case of the 24 hour anneal at 500 $^o$C relative
to the 24 hour anneal at 400 $^o$C, increasing the annealing time
to 96 hours at 500 $^o$C results in a continued contraction of a.
Extending the annealing time an additional order of magnitude
results in a negligible change in a. It appears that continued
annealing at 500 $^o$C causes the structure to become more
disordered in the c- direction while the a- lattice parameter is
returned to near the undamaged value.

\section{Thermodynamic and Transport Measurements}

\begin{figure}
\begin{center}
\includegraphics[angle=0,width=180mm]{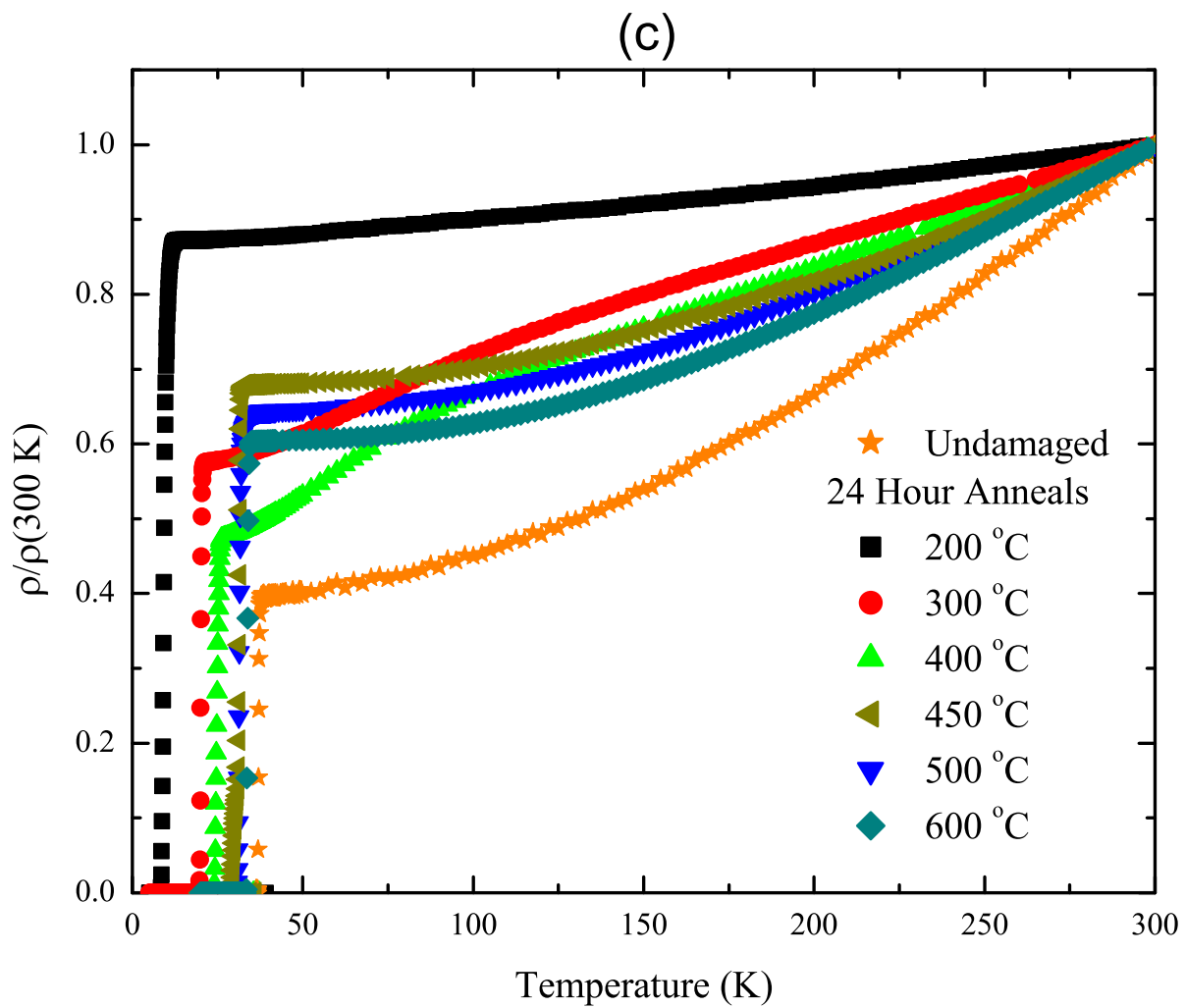}
\end{center}
\caption{Normalized magnetization (a) and zero field resistivity
curves (b) for samples annealed at various temperatures for 24
hours. (c) Normalized resistivity curves for neutron irradiated
Mg(B$_{.962}$C$_{.038}$)$_2$ samples annealed for 24 hour at
various temperatures.}\label{transitions}
\end{figure}

Magnetization and transport measurements were performed to
determine the evolution of T$_c$, H$_{c2}$, and normal state
resistivity as a function of annealing time and temperature. T$_c$
was determined using an onset criteria in resistivity measurements
and a 1\% screening criteria in magnetization. Figure
\ref{transitions}a presents normalized magnetization curves for
the entire set of one day annealed samples. As the annealing
temperature is increased, the transition temperature monotonically
approaches the undamaged value of 36.8 K. It is worth noting that
all of the M(T)) curves show sharp transitions.

Resistivity versus temperature data is plotted in figure
\ref{transitions}b. Normal state resistivity values decrease
monotonically as a function of annealing temperature up until
T$_{anneal}$=450 $^o$C, at which point $\rho_0$ increases
approximately by a factor of four. The exact cause of this jump in
$\rho_0$ is unknown, but it coincides with the broadening of the
(002) x-ray peak. Subsequent increases in the annealing
temperature result in decrease in $\rho_0$ relative to the
T$_{anneal}$=450 $^o$C sample. (It has to be mentioned that an
alternative way of describing these data is to note that the
residual resistivities of samples annealed at 200 $^o$C, 450
$^o$C, 500 $^o$C, and 600 $^o$C are monotonic as a function of the
annealing temperature, raising the question as to whether it is
the resistivities of the 300 $^o$C and 400 $^o$C annealed samples
that are in fact anomalous.) Multiple measurements of samples
annealed at 600 $^o$C for 24 hours are included in figure
\ref{transitions}b and illustrate the spread in the data. It
should be noted that all three of these filaments are from the
same quartz ampoule. All three samples show a decrease in
resistivity relative to that of the 500 $^o$C annealed sample, but
are still above the minimum attained by annealing at 400 $^o$C.

A plot of the normalized, $\rho$/$\rho$(300 K), resistivity shows
that there are two interesting features in the evolution of the
normal state resistivity as a function of annealing temperature
(Figure \ref{transitions}c). For the samples annealed at 300 $^o$C
and 400 $^o$C, the temperature dependence, $\rho$(T), has an odd
hump in the 100 K to 150 K range. Such a hump was not been
observed in pure MgB$_2$ wires \cite{20} but has been reported for
for neutron damaged pure MgB$_2$ annealed up to a temperature of
400 $^o$C for 24 hours \cite{8}. The trends seen in the calculated
normal state resistivity values are also seen in normalized
resistivity (Figure \ref{transitions}c), indicating the apparent
increase in resistivity occurring at 450 $^o$C is a real effect
and not the result of some type of geometrical effect, such as
cracking. It should be noted that whereas the transport
measurements show changes in the evolution of both $\rho_0$ and
$\rho$(T) as a function of annealing temperature for samples
annealed from 300 $^o$C to 450 $^o$C, the magnetization data
showed smooth, sharp transitions with monotonic increases in T$_c$
throughout this range of annealing temperatures.

\begin{figure}
\begin{center}
\includegraphics[angle=0,width=180mm]{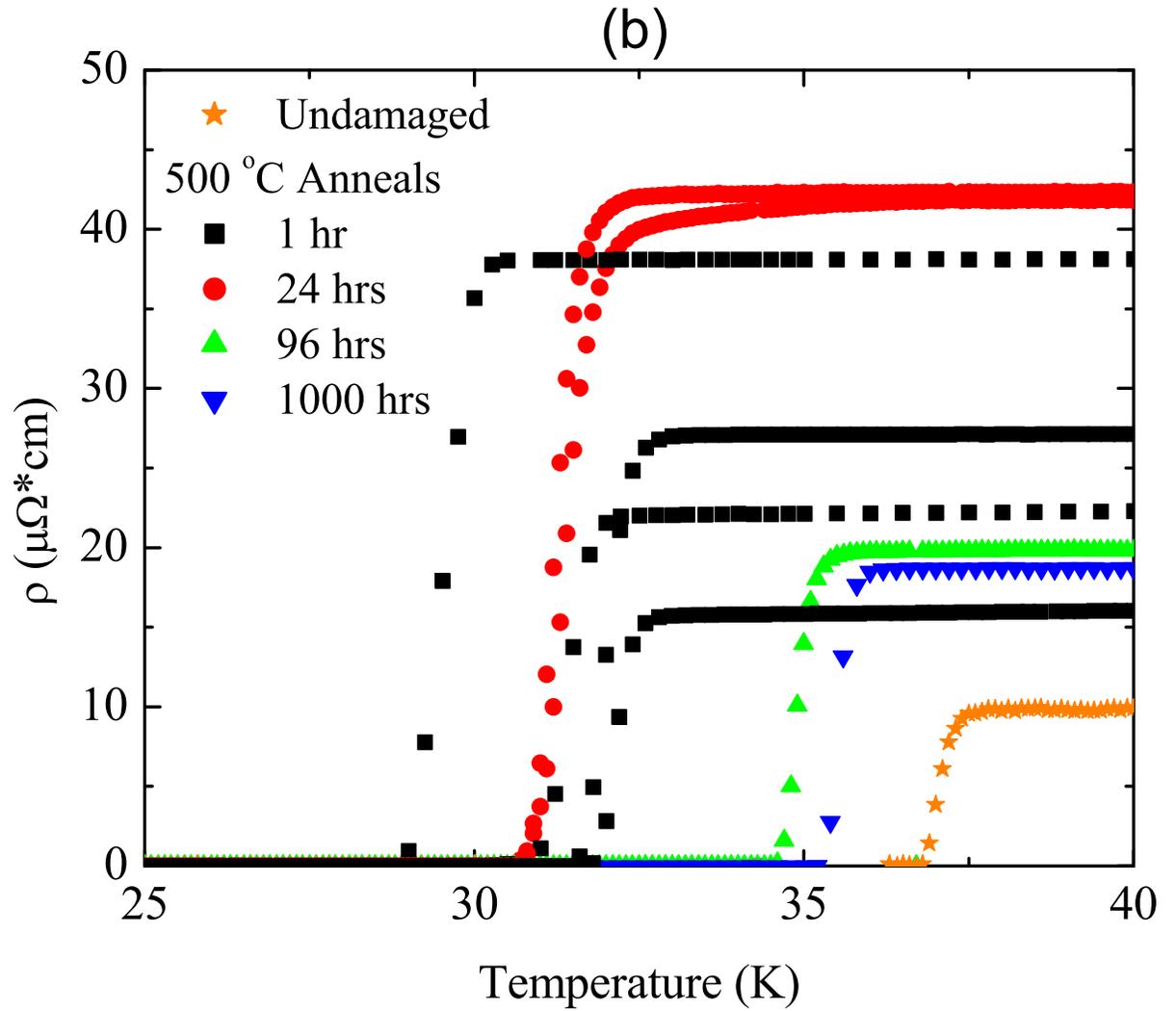}
\end{center}
\caption{Normalized magnetization (a) and zero field resistivity
curves (b) for samples annealed at various times at 500 $^o$C.
}\label{transitions500c}
\end{figure}

Magnetization and transport curves for the series of samples
annealed at 500 $^o$C for various times are plotted in figure
\ref{transitions500c}. Multiple measurements of samples annealed
for 1 hour and 24 hours were performed. The 1 hour anneal
transport data consists of sets of two wires from two different
ampoules. One set was then used for magnetization measurements and
is included along with a filament from a third ampoule. The 24
hour data consists of 2 wires from a single ampoule and an
additional filament from a second ampoule. The normalized
magnetization transitions for the 1 hour anneals all fall below
those of the 24 hour anneals (Figure \ref{transitions500c}a).
However, for the transport data, the range of T$_c$ values is
broader and several of the transitions occur at temperatures
greater than is seen for the 24 hour anneals (Figure
\ref{transitions500c}b). It should be noted that for the samples
annealed at 500 $^o$C for 1 hour for which both magnetization and
transport measurements were performed, in both cases the transport
data showed a T$_c$ that were approximately 0.5 K to 1 K higher
than was observed in the magnetization measurements. Such a spread
was also observed in the case of neutron irradiation on pure
MgB$_2$ \cite{8}, suggesting the spread in the data is a
reflection of sample to sample variation.

The multiple transport measurements on samples annealed at 500
$^o$C for one hour also exhibit considerable spread in the the
normal state resistivity values, and no systematic evolution of
$\rho_0$ as a function of annealing time is observed. Residual
resistivity ratios of these samples indicate that the observed
spread in resistivity values represents true sample to sample
variation in $\rho_0$ and is not the result of pathologic
geometric problems such as cracks.



\begin{figure}
\begin{center}
\includegraphics[angle=0,width=180mm]{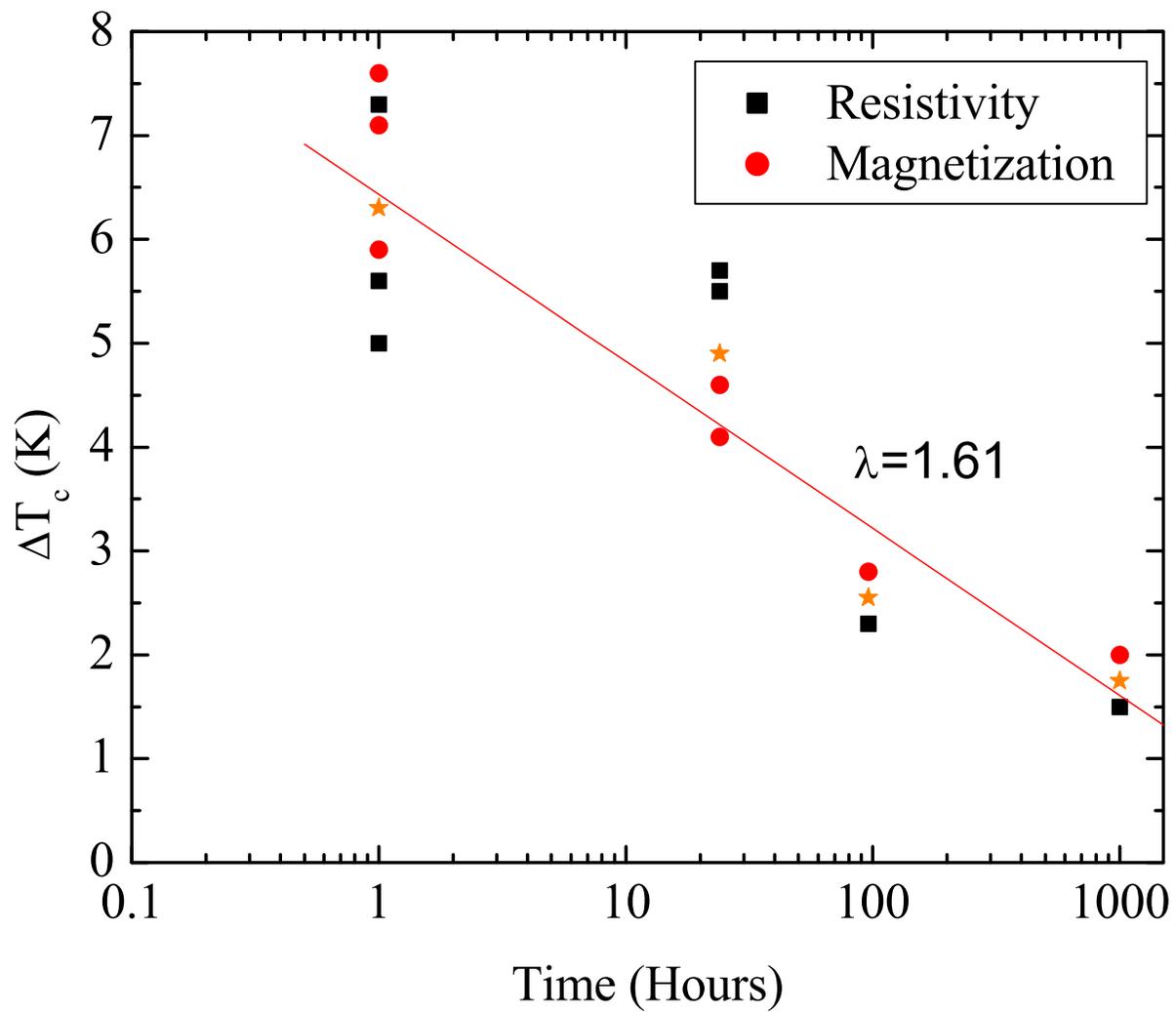}
\end{center}
\caption{$\Delta$T$_c$ as a function of time for 500 $^o$C
anneals. The stars represent the midpoint of spread for a given
time and the rate constant, $\lambda$, is determined by a linear
fit of these data.}\label{tcvtime500c}
\end{figure}

In the case of samples annealed at 500 $^o$C for various times,
generally speaking, extending the annealing time tends to decrease
$\Delta$T$_c$, although there is considerable spread in the data
(Figure \ref{tcvtime500c}). For neutron irradiated pure MgB$_2$
wires, $\Delta$T$_c$, which was taken as a measure of the defect
concentration, was found to decrease exponentially with time
\cite{8}, indicating defects are being annealed out by a single
activated process with random diffusion \cite{18}. For such
samples, the defect concentration, n, obeys the relation

\begin{equation}
n=n_oe^{\lambda t}
\end{equation}

where n$_o$ is the initial defect concentration, $\lambda$ is a
rate constant, and t is the time. $\lambda$ is a function of the
activation energy, E$_a$, and diffusion coefficients. For samples
which do show an exponential behavior in the decrease of
$\Delta$T$_c$ (and by presumption) the defect density as a
function of annealing time the activation energy can be determined
by the so-called cross-cut procedure \cite{18}. This involves
comparing the annealing time for which different temperature
anneals reached the same defect density, i.e. $\Delta$T$_c$. If a
$\Delta$T$_c$ is reached by annealing at a temperature T$_1$ for a
time t$_1$ and by annealing at a temperature T$_2$ for a time
t$_2$, then the activation energy is related to these quantities
by:

\begin{equation}
ln \frac{t_1}{t_2} = \frac{E}{k} (\frac{1}{T_1} - \frac{1}{T_2}).
\end{equation}

Unfortunately, we did not have a sufficient number of samples to
perform studies of the temperature dependance of $\Delta$T$_c$ as
a function of annealing time at any temperatures other than 500
$^o$C. However, if we assume that the primary difference between
neutron irradiation of pure and carbon doped MgB$_2$ is in the
activation energies, then we can gain some insight by comparing
$\lambda$ values attained from the two sets of samples. For the
case of the carbon doped samples, since the spread in T$_c$ for a
given annealing time is so large, we can obtain only a rough
estimate, based on a fitting of the midpoint of the T$_c$ values.
Such a calculation yields a rate constant of $\lambda$=1.61
s$^{-1}$. For irradiation on pure MgB$_2$ anneals at 200 $^o$C,
300 $^o$C, and 400 $^o$C yielded rate constants of 2.77 s$^{-1}$,
2.21 s$^{-1}$, and 3.51 s$^{-1}$ respectively \cite{8}. Although
there is considerable spread in the rate constants for the pure
MgB$_2$ case, and considerable uncertainty in the rate constant
associated with the carbon doped MgB$_2$ samples (estimates of
$\lambda$ from raw data range from 1.00 s$^{-1}$ to 2.03 s$^{-1}$)
analysis of equation 2 shows that since carbon doped samples have
a lower rate constant than pure MgB$_2$ samples the activation
energy is higher for carbon doped samples. It should be noted that
higher reaction temperatures were required to synthesize carbon
doped MgB$_2$ than pure MgB$_2$ \cite{9}, suggesting that there
are different energy scales associated with carbon doping.

\begin{figure}
\begin{center}
\includegraphics[angle=0,width=180mm]{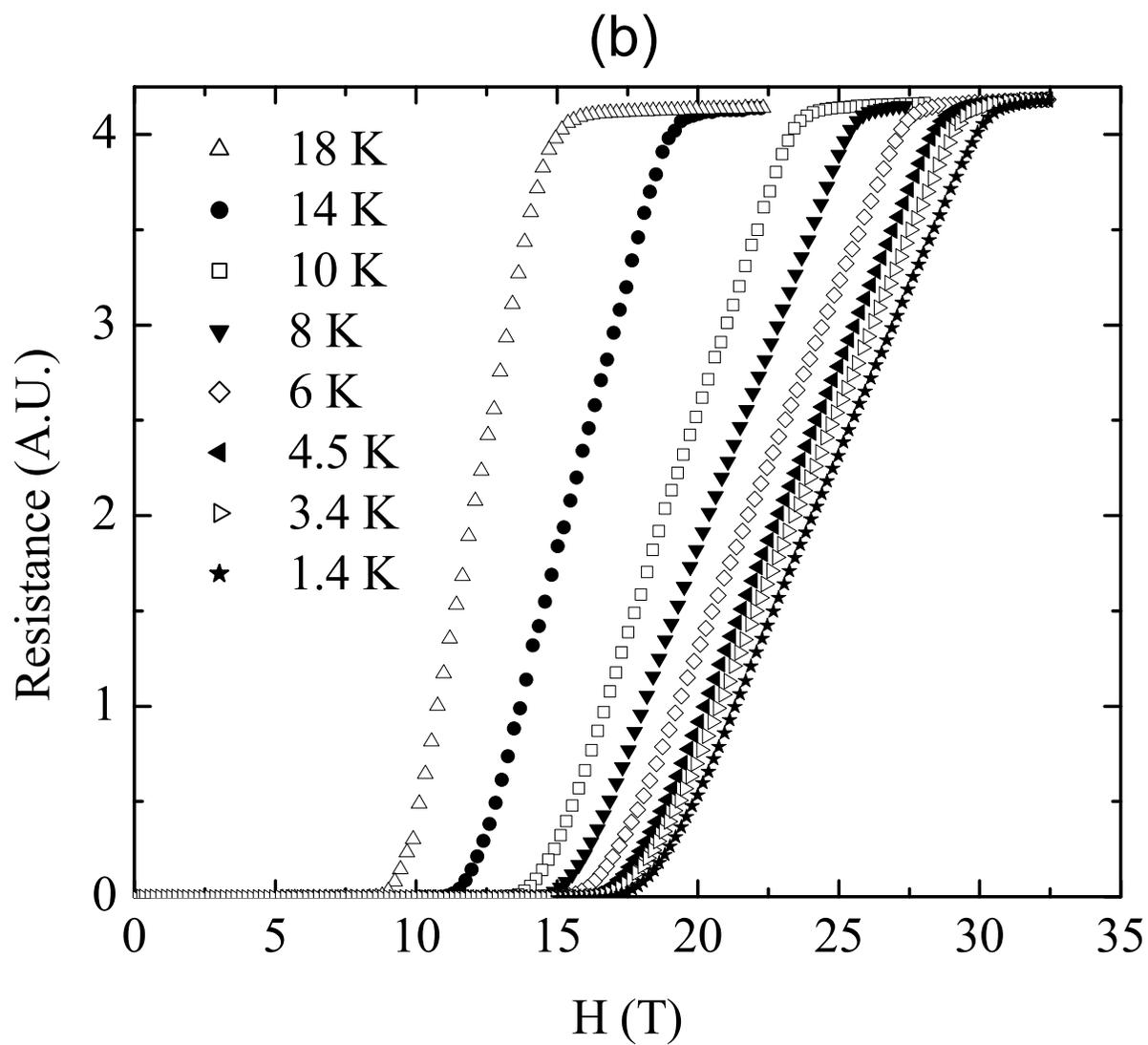}
\end{center}
\caption{Resistivity versus temperature (a) and resistivity versus
field (b) for a sample annealed at 500 $^o$C for 1000
hours.}\label{infieldrho}
\end{figure}

\begin{figure}
\begin{center}
\includegraphics[angle=0,width=180mm]{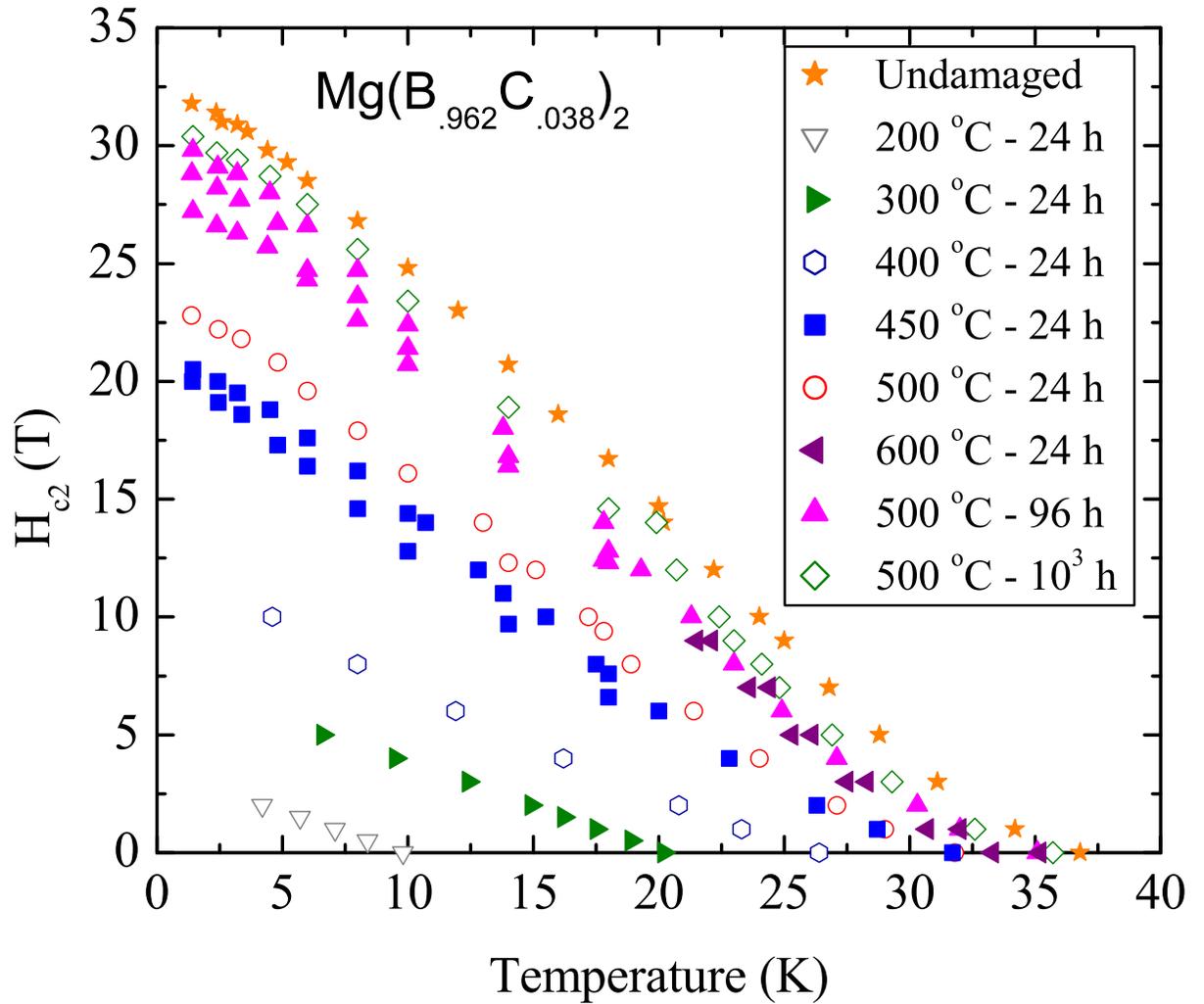}
\end{center}
\caption{Upper critical field curves for samples annealed for 24
hours at various temperatures and samples annealed for various
times at 500 $^o$C. Multiple data sets exist for the 450 $^o$C/24
hour and 500 $^o$C/96 hour anneals. H$_{c2}$(T=0) approximately
scales with T$_c$.}\label{hc2plot}
\end{figure}

Upper critical field values were determined using the onset
criteria in both resistivity versus temperature in applied fields
up to 14 T and resistivity versus field in field sweeps up to 32.5
T (Figure \ref{infieldrho}). The complete H$_{c2}$(T) curves for
the entire set of annealing profiles is given in figure
\ref{hc2plot}. The set forms a sort of "Russian doll pattern,"
with H$_{c2}$(T=0) approximately scaling with T$_c$. Such behavior
was also observed for neutron irradiation and post-exposure
anneals on pure MgB$_2$ filaments \cite{8}.

\begin{figure}
\begin{center}
\includegraphics[angle=0,width=180mm]{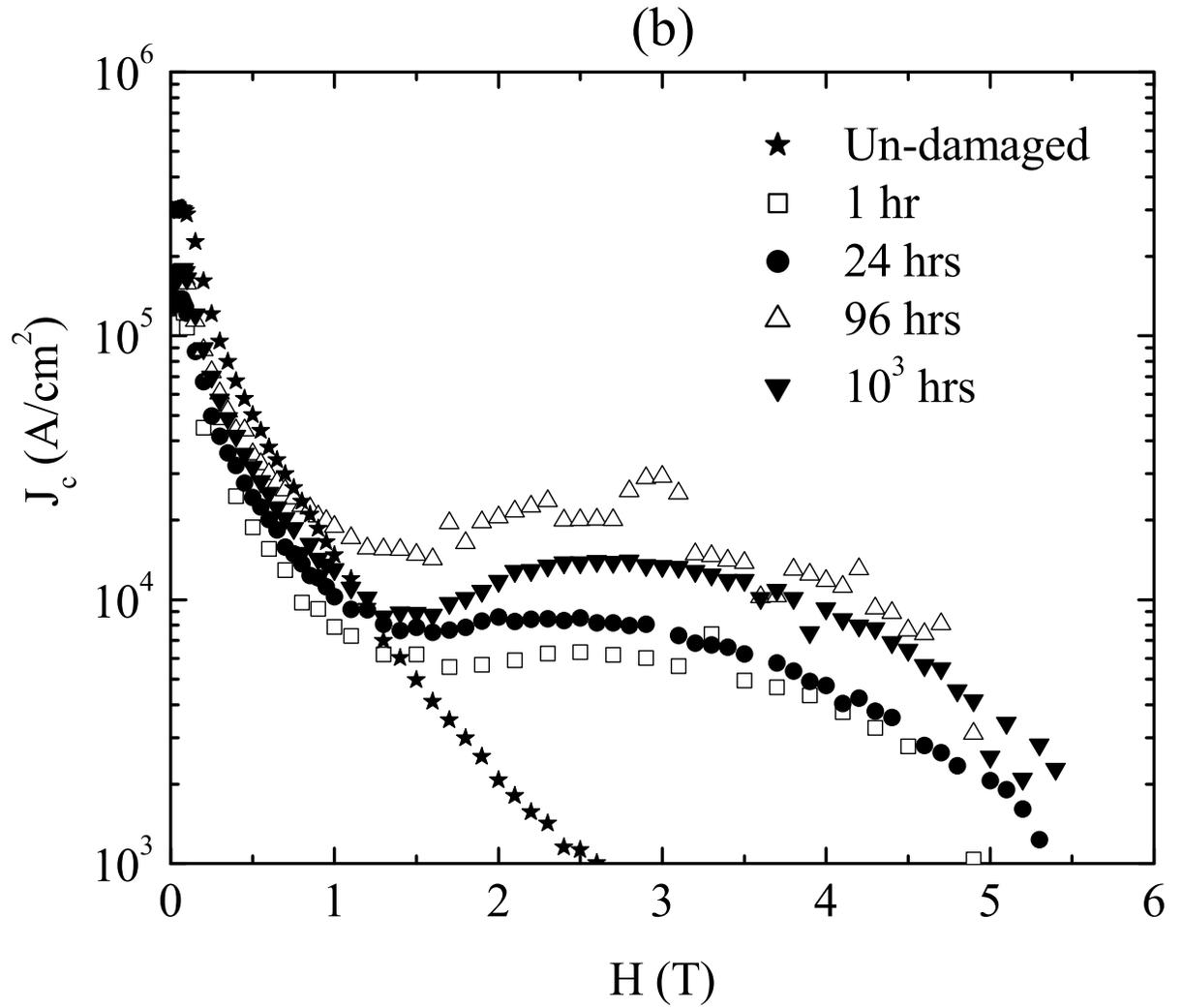}
\end{center}
\caption{Critical current densities at 5 K inferred from
magnetization hysteresis loops. (a) Samples annealed at various
temperatures for 24 hours. (b) Samples annealed at 500 $^o$C for
various times.}\label{jcplots}
\end{figure}

Critical currents densities were estimated from magnetization
hysteresis loops using the Bean Critical State Model \cite{12} for
cylindrical geometry. Figure \ref{jcplots}a shows the field
dependence of J$_c$ at T=5 K for the entire set of 24 hour anneals
at various annealing temperatures. For the set of one day anneals,
the in field performance improves with annealing temperature, up
to 500 $^o$C. Presumably the increase transition temperature and
hence decreased reduced temperature (T/T$_c$) of the measurement,
is the primary cause of the enhancement. However, their exists a
much greater increase in in-field J$_c$ values between the sample
annealed at 400 $^o$C and the sample annealed at 450 $^o$C than
between the sample annealed at 300 $^o$C and the sample annealed
at 400 $^o$C. Whereas increasing the annealing temperature from
300 $^o$C to 400 $^o$C results in a near 7 K or 36\% increase in
T$_c$ but only a factor of two improvement in J$_c$ at all fields,
increasing the annealing temperature from 400 $^o$C to 450 $^o$C
results in only an additional 5 K increase or 17\% increase in
T$_c$, but nearly an order of magnitude increase in in-field J$_c$
values. This suggests the increase in disorder between the
hexagonal planes, which coincided with in increase in resistivity,
may also play a role in enhancing the flux pinning. The sample
annealed at 600 $^o$C shows slightly reduced J$_c$(H) values
relative to the 500 $^o$C anneal. Thus, at 600 $^o$C, either the
increased order along c is diminishing J$_c$, or we are beginning
to anneal out some of the defects which are effective at pinning
vortices, or some combination thereof.

Extending the annealing time from 1 hour to 96 hours at 500 $^o$C
results in relatively minor increases in in-field J$_c$ values
(Figure \ref{jcplots}b). The data for the 500 $^o$C and 96 hour
anneal is particularly noisy and may not truly be enhanced
relative to the 500 $^o$C/10$^3$ hour anneal. All of these 500
$^o$C anneals show J$_c$ remaining fairly constant over the field
range of 1-4 Tesla.

At 5 K, the best in field performance, which results from the 96
and 1000 hour anneals at 500 $^o$C, yield J$_c$ value which
maintain 10$^4$ A/cm$^2$ in an applied field of 4 T. In
comparison, the best in field performance at 5 K for a sample of
neutron irradiated pure MgB$_2$ dropped below 10$^4$ A/cm$^2$ in
an applied field of approximately 1.5 T \cite{8} and an undamaged
carbon doped filament dropped below this level near 1 T \cite{9}.

\section{Discussion}

Up to an annealing temperature of 400 $^o$C, the behavior of
irradiated carbon doped samples is quite similar to that of
irradiation of pure MgB$_2$ filaments. As-damaged samples have
expanded unit cells and suppressed superconducting transition
temperatures. Post exposure annealing tends to return the lattice
parameters and T$_c$ towards the undamaged values. In both cases,
the upper critical field values, H$_{c2}$(T=0) tend to scale with
T$_c$.

The intriguing aspect of neutron damage in the carbon doped
samples is the temperature induced decrease in structural order in
the c- direction which occurs near T$_{anneal}$=450 $^o$C. No such
effect was seen in the case of neutron irradiation in pure MgB$_2$
filaments. The feature doesn't appear to be a phase segregation as
no double transition is seen in magnetization. Furthermore,
samples annealed at 500 $^o$C showed sharp, single transitions in
both magnetization and resistivity, while the having broad (002)
peaks. The sample annealed for 24 hours at 600 $^o$C showed
evidence of having two distinct phases with different c- lattice
parameters but the same a- lattice parameter. Here again, both the
magnetization and transport data showed only single transitions.
Thus the transition temperature appears to be insensitive to the
c- lattice parameter. The a- lattice parameter is presumably not
the sole determining factor for T$_c$, as samples annealed for 24
hours at 450 $^o$C and 500 $^o$C showed increases in $\Delta$a
relative to the sample annealed for 24 hours 400 $^o$C, while also
having increased T$_c$ values. It should be noted that although
measurements on single filaments showed sharp, single transitions,
there was significant spread in the T$_c$ values between different
samples.

The broadening of the (002) peak coincides with a near 4 fold
increase in the normal state resistivity and approximate order of
magnitude increase in in-field J$_c$ values. Increases in normal
state resistivity for annealing temperatures above 400 $^o$C have
been observed in neutron irradiation of pure MgB$_2$ \cite{10}. In
this case, the authors report a continued increase in $\rho$$_0$
as a function of annealing temperature above 400 $^o$C and
attribute it to a loss of intergrain connectivity. In contrast our
data show a decrease in $\rho_0$ for 24 hour anneals at 500 $^o$C
and 600 $^o$C, which supports the notion that a different
mechanism is responsible for the anomalous behavior in the normal
state resistivity of neutron irradiated carbon doped samples.

Despite the decrease in structural order along the c- axis and
large increase in resistivity at T$_{anneal}$=450 $^o$C, no
qualitative changes to the evolution of T$_c$ and H$_{c2}$ as a
function of annealing temperature were observed. The transition
temperature tends to approach that of the undamaged sample and
H$_{c2}$ continues to roughly scale with T$_c$. Even the 500
$^o$C, 1000 hour anneal shows a suppressed H$_{c2}$(T=0) value
relative to the undamaged sample. This is in contrast to the
results obtained with irradiation of pure MgB$_2$ filaments
\cite{8}. For fluence levels of 4.75*10$^{18}$ n/cm$^2$ and
9.50*10$^{18}$ n/cm$^2$, two values which envelop the
7.13$*$10$^{18}$ n/cm$^2$ fluence level used in this study,
H$_{c2}$(T=0) was enhanced by approximately 2 T and 3 T
respectively.

\begin{figure}
\begin{center}
\includegraphics[angle=0,width=180mm]{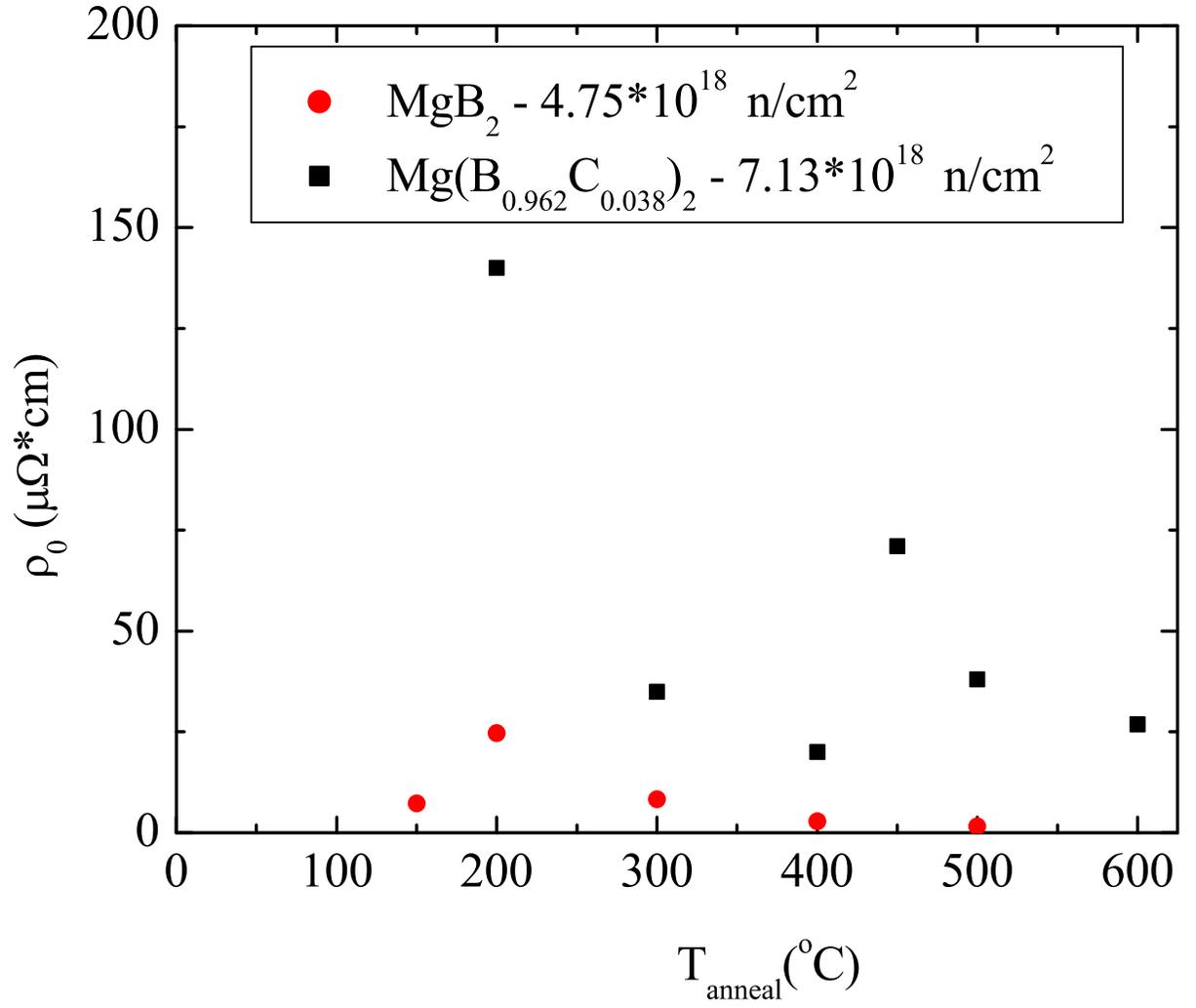}
\end{center}
\caption{Evolution of the normal state resistivity as a function
of annealing temperature in neutron irradiated pure and carbon
doped MgB$_2$ filaments.}\label{rhoVtanneal}
\end{figure}

Both pure and carbon doped samples show increases in normal state
resistivity after annealing (Figure \ref{rhoVtanneal}). Pure
MgB$_2$ wires irradiated with a fluence of 4.75*10$^{18}$ n/cm$^2$
and subsequently annealed for 24 hours at various temperatures
showed an increase in $\rho_0$ at an annealing temperature of 200
$^o$C. In contrast, for Mg(B$_{.962}$C$_{.038}$)$_2$ filaments
exposed to a fluence of 7.13*10$^{18}$ n/cm$^2$ and subsequently
annealed for 24 hours at various temperatures, the apparent
anomalous increase occurs for an annealing temperature of 450
$^o$C. For both sets of samples, following the abrupt increase,
the normal state resistivity decreases monotonically as a function
of annealing temperature. It should be noted however, that, in the
case of the carbon doped samples, it is not clear whether or not
there exists some correlation between the change in the
temperature dependence of the normal state resistivity for the 300
$^o$C and 400 $^o$C anneals and the reduced $\rho_0$ values for
these anneals. That is, rather than the jump in $\rho_0$ at 450
$^o$C representing an anomalous increase, the $\rho_0$ values
associated with the samples annealed at 300 $^o$C and 400 $^o$C
may in fact be anomalously low.

These data, coupled with the analysis of the rate constants,
suggest that in the case of neutron irradiated carbon doped
MgB$_2$ samples, there is a different, and higher, energy scale
associated with the annealing of defects. This is perhaps not
altogether surprising, as it was shown that higher reaction
temperatures were necessary to form the carbon doped phase.
Whereas pure boron filaments can be fully converted to MgB$_2$ in
as little as 2 hours at 950 $^o$C \cite{19}, isothermal reactions
required 48 hours at 1200 $^o$C to convert carbon doped boron
fibers to Mg(B$_{1-x}$C$_x$)$_2$, even for doping levels as low as
x=0.004 \cite{9}. Although diffusion of Mg vapor into a carbon
doped boron matrix and repair of neutron induced damage are indeed
two, rather different, phenomena, this correlation is at least
worth noting.

The temperature dependence of the upper critical field shows
positive curvature near T$_c$ for samples with a T$_c$ near 26 K
and above (Figure \ref{hc2plot}). For the samples annealed for 24
hours at 200 $^o$C and 300 $^o$C H$_{c2}$ approaches T$_c$
linearly. If the two bands are fully mixed, the temperature
dependance of the upper critical field should follow WHH \cite{13}
behavior, where H$_{c2}$(T=0) is given by:

\begin{equation}
H_{c2}(T=0)=0.69 T_c \frac{dH_{c2}}{dT}.
\end{equation}

Calculating H$_{c2}$(T=0) using equation 3 for the samples
annealed for 24 hours at 200 $^o$C, 300 $^o$C, and 400 $^o$C
yields values of 2.4 T, 5.2 T, and 6.5 T. Experimentally, lowest
temperate H$_{c2}$ values we could reliably attain from transport
measurements for each of these samples were 2.1 T, 6.5 T, and 10
T, occurring at temperatures of 4.2 K, 5.2 K, and 4.6 K
respectively. Only for the sample annealed at 200 $^o$C does the
WHH fit yield an H$_{c2}$(T=0) value which is consistent with the
experimentally observed data. Thus it is likely that only for
samples with a T$_c$ below 10 K are the bands fully mixed. Such a
result was also obtained in the case of the neutron irradiation on
pure MgB$_2$ \cite{8}. This implies that the inter-band scattering
rates resulting from defects associated with the neutron
irradiation are comparable in both pure and carbon doped samples,
suggesting that the scattering associated with the two different
sources of defects act independently. This is not altogether
surprising as the suppression in T$_c$ for carbon doped samples is
believed to result from changes in the Fermi surface, rather than
increases in inter-band scattering and scattering associated with
carbon doping is believed to be primarily within the $\pi$ band
\cite{3}. Thus if inter-band scatting is introduced primarily
through defects resulting from neutron damage, if pure and carbon
doped samples have similar initial T$_c$ values, then they should
also exhibit fully mixed bands at similar T$_c$s, which was
experimentally observed. A similar result was found in the case of
neutron irradiation on pure MgB$_2$ filaments, implying that the
scattering associated with carbon doping and neutron irradiation
act independently.

\section{Conclusions}

We have studied superconducting and normal state properties of
neutron irradiated carbon doped MgB$_2$ filaments as a function of
post exposure annealing time and temperature. In spite of
anomolous behavior in the evolution of the c- lattice parameter
and normal state resistivity T$_c$ tended to return towards the
undamaged value with increased annealing time and temperature.
Upper critical field values were found to approximately scale with
T$_c$ and exhibited WHH like behavior, suggesting a complete
mixing of the two bands, for samples with T$_c$ near 10 K. Neutron
irradiation of pure MgB$_2$ also led to a complete mixing of the
bands near 10 K, suggesting that the scattering associated with
carbon doping and neutron irradiation act independently.

\textbf{Acknowledgements}

Ames Laboratory is operated for the U.S. Department of Energy by
Iowa State University under Contract No. W-7405-Eng-82. This work
was supported by the Director for Energy Research, Office of Basic
Energy Sciences. A portion of this work was performed at the
National High Magnetic Field Laboratory, which is supported by NSF
Cooperative Agreement No. DMR-0084173 and by the State of Florida.

$^\dagger$In the course of this work Raymond J. Suplinskas
succumbed to a long term illness. Ray's enthusiastic support and
collaboration was vital to much of the success of our MgB$_2$
research. We will miss him greatly.

\clearpage

\clearpage

\clearpage

\clearpage

\begin{figure}
\begin{center}
\includegraphics[angle=0,width=180mm]{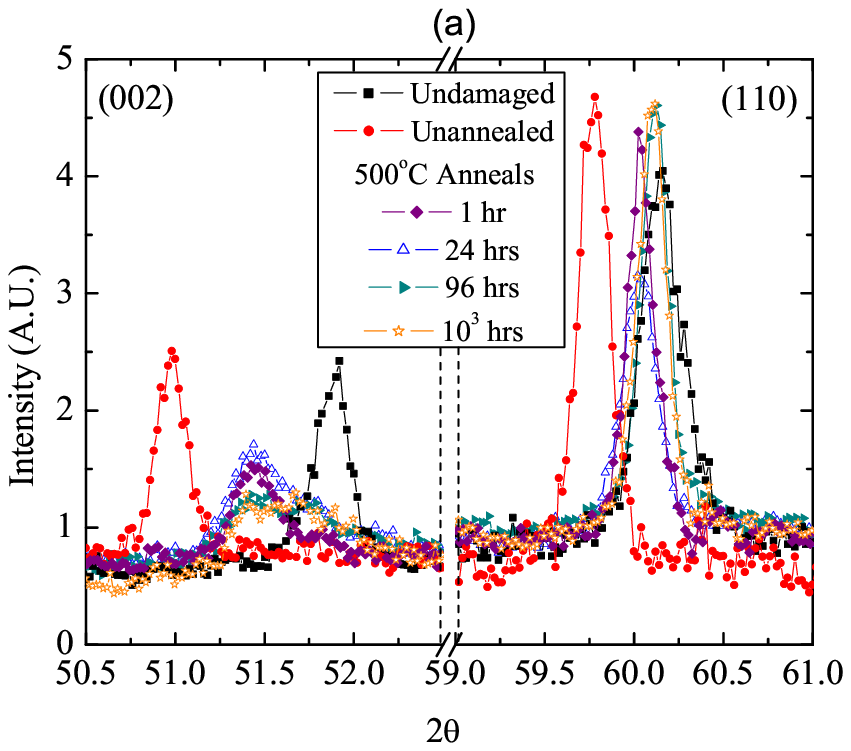}
\end{center}
\end{figure}

\clearpage

\clearpage

\begin{figure}

\begin{center}
\includegraphics[angle=0,width=180mm]{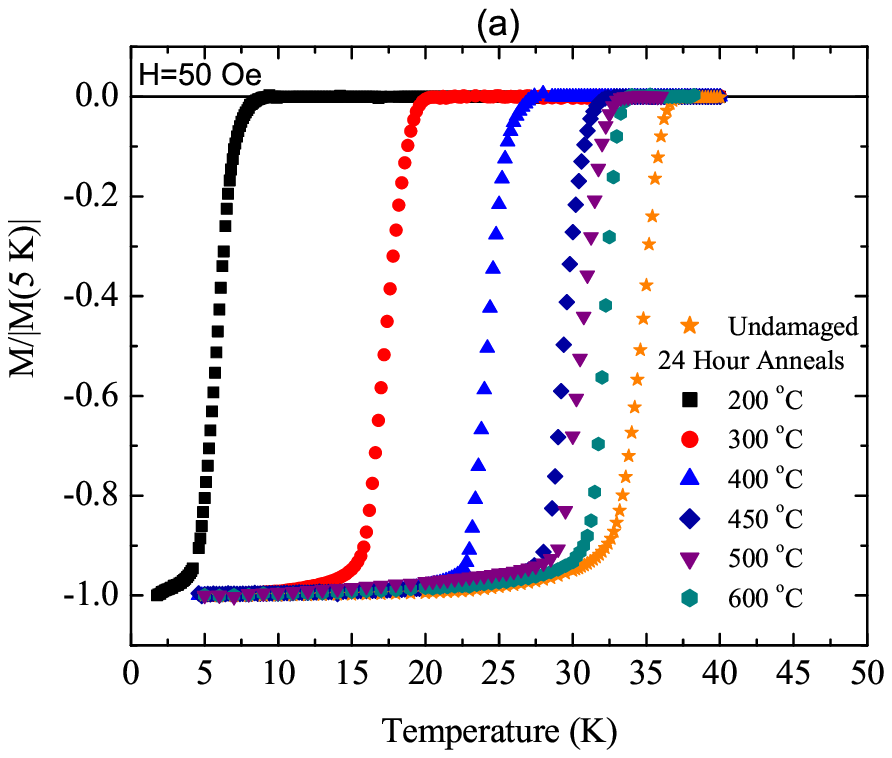}
\end{center}
\end{figure}

\clearpage

\begin{figure}
\begin{center}
\includegraphics[angle=0,width=180mm]{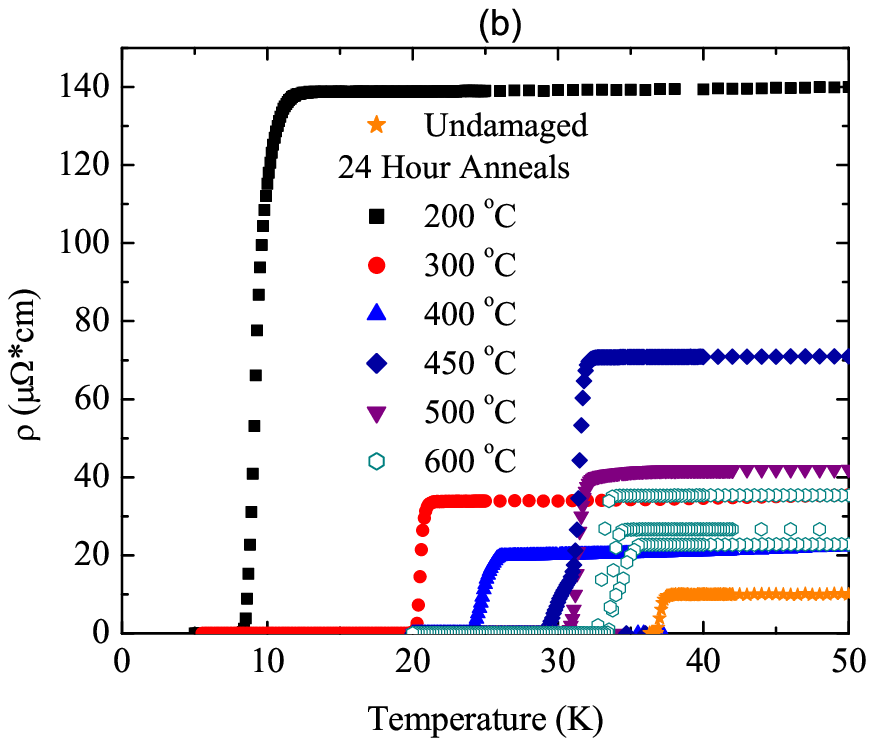}
\end{center}
\end{figure}

\clearpage

\clearpage

\begin{figure}
\begin{center}
\includegraphics[angle=0,width=180mm]{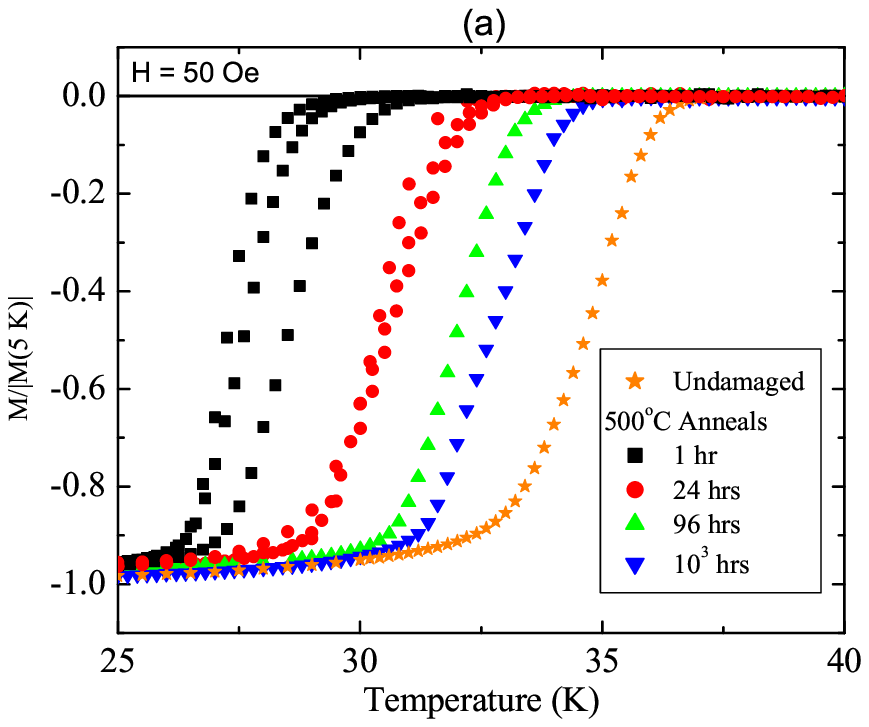}
\end{center}
\end{figure}

\clearpage

\clearpage

\clearpage

\begin{figure}
\begin{center}
\includegraphics[angle=0,width=180mm]{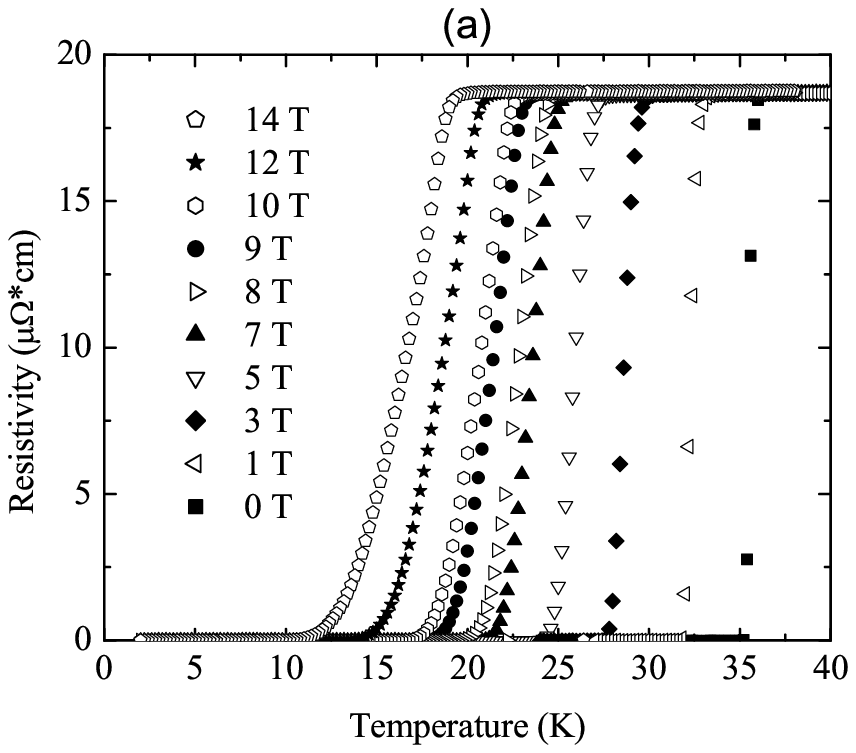}
\end{center}
\end{figure}

\clearpage

\clearpage

\clearpage

\begin{figure}
\begin{center}
\includegraphics[angle=0,width=180mm]{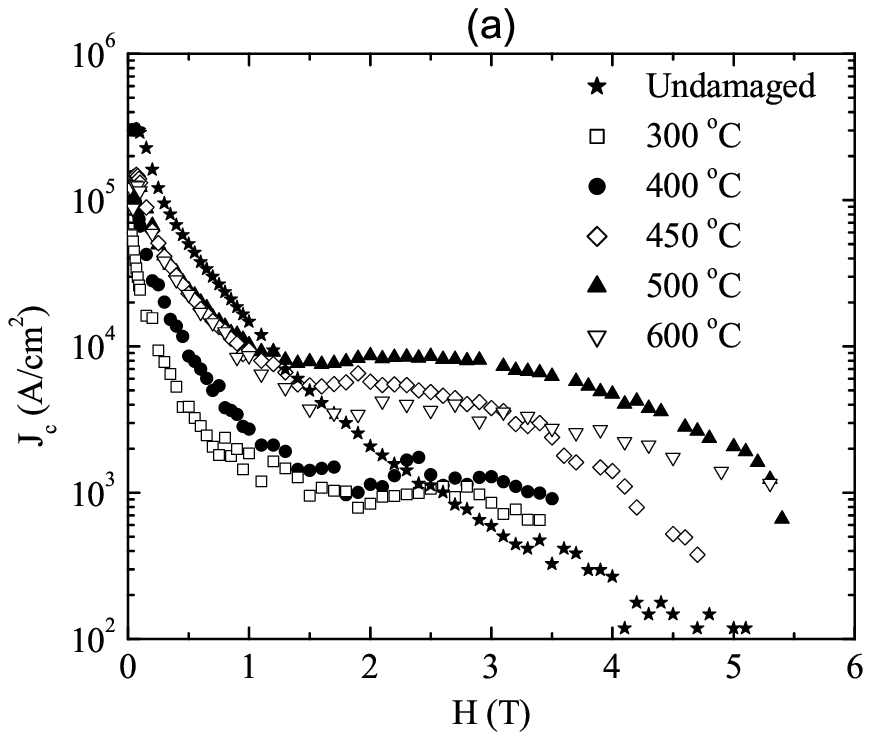}
\end{center}
\end{figure}

\clearpage

\clearpage

\end{document}